# CEMP: a platform unifying high-throughput online calculation, databases and predictive models for clean energy materials

*Jifeng Wang[1], Jiazhe Ju[1], Ying Wang[1]\**

1. State Key Laboratory of Molecular Engineering of Polymers
   Department of Macromolecular Science
   Fudan University
   AI for Polymer Science Research Center
   Shanghai, 200438, China

**Abstract**: The development of materials science is undergoing a shift from empirical approaches to data-driven and algorithm-oriented research paradigm. The state-of-the-art platforms are confined to inorganic crystals, with limited chemical space, sparse experimental data and a lack of integrated online computation for rapid validation. Here, we introduce the Clean Energy Materials Platform (CEMP), an open-access platform that integrates high-throughput computing workflows, multi-scale machine learning (ML) models and a comprehensive materials database tailored for clean energy applications. A key feature of CEMP is the online calculation module, which enables fully automatic quantum and molecular dynamics simulations via structured table uploads. CEMP harmonizes heterogeneous data from experimental measurements, theoretical calculation and AI-based predictions for four material classes, including small molecules, polymers, ionic liquids, and crystals. The platform hosts ~ 376,000 entries, including ~6,000 experimental records, ~50,000 quantum-chemical calculations and ~320,000 AI-predicted properties. The database covers 12 critical properties and the corresponding ML models demonstrate robust predictive power with $R^2$ ranging from 0.64 to 0.94, thus ensures rapid material screening, structure-property relationship analysis and multi-objective optimization for clean energy applications. CEMP aims to establish a digital ecosystem for clean energy materials, enabling a closed-loop workflow from data acquisition to material discovery and real-time online validation.

## INTRODUCTION

The advancement of next-generation batteries and renewable energy systems critically depends on the timely discovery of high-performance materials. However, conventional trial-and-error methods remain slow, labor-intensive, and inefficient. [1,2] To address this bottleneck, the field is witnessing a fundamental shift toward computationally accelerated discovery, leveraging theoretical calculations, high-throughput screening and machine learning (ML).[3,4] These approaches have shown considerable promise in identifying functional materials with both efficiency and accuracy in practical applications. [5,6]

To meet this demand, several materials design platforms have emerged[7-15]. As systematic comparison shown in **Figure 1**, these platforms commonly support structural visualization, batch data retrieval and online property prediction, and have proven effective for screening inorganic solids. Despite recent progress, existing platforms face key limitations. First, most platforms are confined to inorganic crystals without emphasizing other material systems, including polymers, small molecules, and ionic liquids, thus exhibits limited cross-system screening. Second, they typically offer static workflows with limited support for user-defined structures, interactive analysis and real-time calculations in exploring novel materials. Third, most platforms rely heavily on computational databases with limited integration of experimental measurements. For many key properties, such as conductivity or mechanical properties, experimental validation is essential, and predictions based solely on theoretical data may fail to reflect real-world performance.

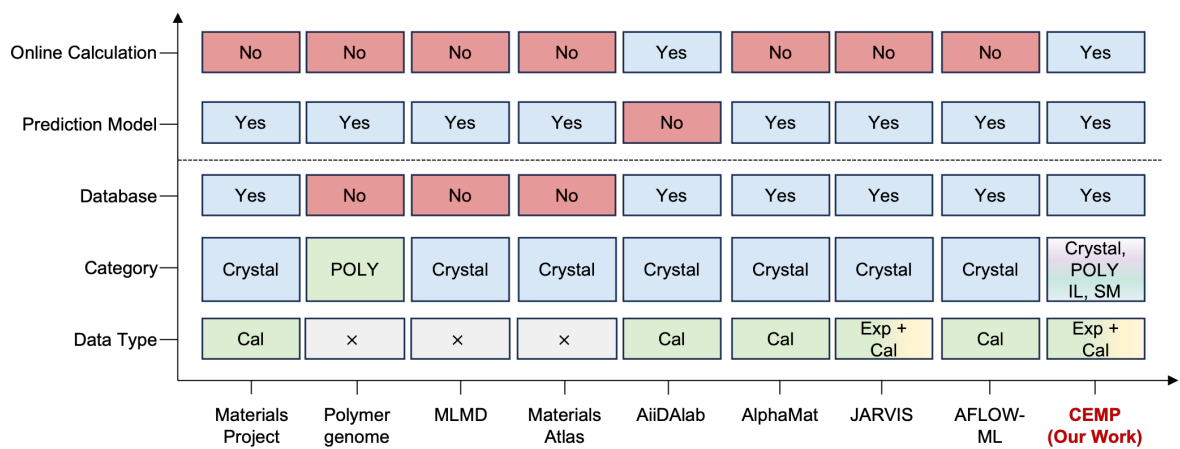

**Figure 1.** Cross-platform comparison of materials design infrastructures based on five key criteria.[7-15]。 The comparison includes: (1) Online Calculation – whether the platform supports

first-principles, molecular dynamics simulations, or quantum chemical calculations directly online; (2) Prediction Model – availability of property prediction models; (3) Database – integration of a built-in materials database; (4) Category – types of materials supported, including crystals (Crystal), polymers (POLY), ionic liquids (IL), and small molecules (SM); (5) Data Type – sources of data within the database, where "Cal" refers to computed data (e.g., from theoretical calculations), and "Exp" denotes experimentally measured values.

To overcome the mentioned obstacles, real-time online computation is increasingly recognized as a critical capability.[16] Compared with the black-box predictions from static models, online simulations offer interpretable, trustworthy results. It allows flexibly in defining structures and parameters to validate predictions via physics-based computations and enhance generalizability across unexplored chemical spaces. Here, we introduce the Clean Energy Materials Platform (CEMP)—an open-source, integrated infrastructure that unifies high-throughput computation, materials databases, and AI models for clean energy applications. CEMP enables automated online workflows for quantum chemistry (QC) and molecular dynamics (MD) simulations through simple spreadsheet uploads. The system manages structure generation, job scheduling, error handling, and output parsing, offering complete calculation files for user access and significantly lowering the barrier to high-performance computing. In parallel, the platform integrates a multi-type materials database encompassing polymers, small molecules, ionic liquids, and crystals, comprising ~376,000 structure–property records sourced from peer-reviewed literature, QC calculations, and ML pipelines. This comprehensive dataset supports robust model development and cross-domain screening. Building on this foundation, CEMP hosts a suite of predictive models tailored to different material classes, capable of estimating essential properties such as electrochemical stability, mechanical properties and ionic conductivity. Furthermore, higher-level models extend predictive capabilities to device-relevant metrics, including battery cycling performance, thereby bridging the gap between molecular design and real-world application.

By establishing the unified, scalable, and automated "structure–prediction–computation" pipeline, CEMP accelerates clean energy materials screening and provides a next-generation digital infrastructure innovation for cross-domain and cross-scale material development.

## PLATFORM ARCHITECTURE

CEMP is built upon the Django framework, using Python and standard web technologies (e.g., HTML, javascript et. al) and features a robust back-end infrastructure coupled with an intuitive user interface. It offers a unified and visualized gateway for materials modeling and analysis. By adopting a form-driven and task-oriented workflow, CEMP simplifies complex theoretical calculations into one-click operations, significantly lowering the barrier for non-expert users. Users are only required to select the desired calculation type, complete a parameter template and upload it to the platform. CEMP then automatically dispatches the job to high-performance computing resources and returns the calculated properties along with the generated structure files (**Figure 2**). The platform is organized into five core modules, each tailored to a specific class of materials and collectively supporting the end-to-end design of clean energy systems:

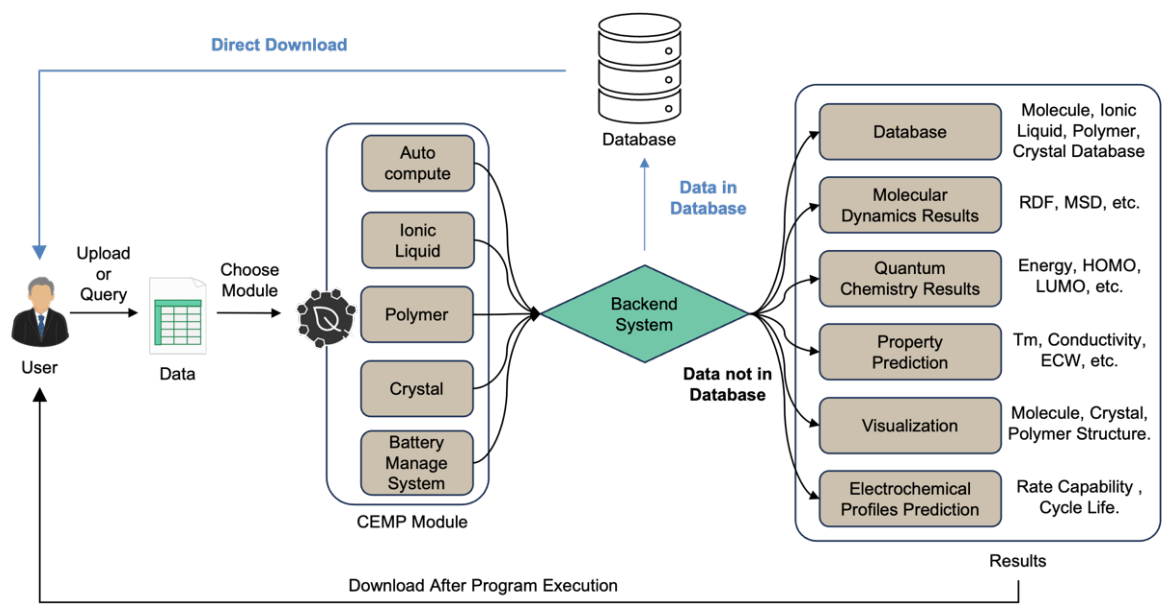

**Figure 2**. Overview of the CEMP platform architecture. CEMP is organized into five core modules, including Autocompute, Ionic Liquid, Polymer, Crystal, and Battery Management System. The platform supports data retrieval, structure visualization, property prediction, electrochemical curve forecasting, and automated execution of QC calculations and MD simulations. An internal computation log system is embedded to prevent redundant tasks and improve overall efficiency.

**Autocompute Module:** Serving as the computational backbone of CEMP, the Autocompute module integrates automated high-throughput workflows for both QC calculations and MD simulations. It supports the full process from structure generation to property analysis. A built-in small molecule database containing ~20,000 data from QC calculation is included, along with a computation history tracker that avoids redundant tasks and improves efficiency.

**Ionic Liquid Module:** This module is specifically designed for ionic liquid screening, integrating ~110,000 entries from experimental, QC, and ML data. It supports property prediction (e.g., melting point, ionic conductivity, electrochemical window) and structure-based filtering, facilitating the discovery of high-performance ionic liquids.

**Polymer Module:** For polymeric materials, this module includes a curated database of ~230,000 records spanning experimental, QC and ML data. It provides predictive models for key properties such as glass transition temperature, dielectric constant, melting point, tensile strength, and Young's modulus. Additional tools support polymer structure generation and topology file construction, enabling full-chain design and analysis for polymer performance optimization.

**Crystal Module:** To evaluate crystalline materials for battery electrodes, this module integrates a crystal structure database (~10,000 entries) and predictive models for electrochemical performance, including average voltage, specific capacity, and energy density. It also offers interactive 3D visualization of crystal structures for enhanced interpretability.

**Battery Management System Module:** This module hosts an experimental-data-driven database for battery electrochemical performance, with a focus on rate capability and cycling stability. It enables the development of battery performance evolution models, bridging the gap between material-level predictions and device-level behaviors.

In summary, CEMP adopts a modular, unified architecture that spans the full design pipeline for clean energy materials from molecular structures to device-level applications. Each module focuses on a distinct material class and integrates dedicated databases, automated simulation workflows and tailored predictive models. This design enables seamless progression from structure generation to property evaluation and application validation, providing a scalable and interoperable platform for cross-domain materials discovery.

**ONLINE HIGH-THROUGHPUT COMPUTATIONAL WORKFLOW**

To address the limitations of conventional ML models, including narrow applicability, limited extrapolation ability and a lack of physical interpretability, the CEMP platform integrates an online high-throughput computational workflow combining QC calculation and MD simulation. This workflow generates traceable, physically meaningful data, especially valuable in chemical spaces with limited experimental coverage. It supports on-demand simulations of user-defined systems, enabling flexible exploration of novel or complex materials beyond existing databases. In addition, it provides direct access to microscopic properties, such as energies, diffusion coefficients, and structural correlations.

The workflow operates through five functional layers: input, preprocessing, scheduling, computation, and results output (**Figure 3**). A summary of task types and key output properties is provided in **Table 1**, with additional computational capabilities planned for integration in future updates. To improve accessibility, CEMP provides a spreadsheet-based submission interface in which users upload a table containing SMILES strings and task parameters. The system then automatically manages the entire workflow, including job configuration, simulation execution and results delivery.

**Table 1.** Summary of task types and computed properties in the high-throughput computational workflow.

| Workflow | Task Type | Output Properties |
|---|---|---|
| Quantum Chemistry Calculation | Monomer Energy | Energy, HOMO, LUMO, Dipole, … |
| | Structure Optimization | Optimized Structure, H, G, S, … |
| | Dimer Energy | Binding Energy, Dimer Energy, Component Energy… |
| | Redox Potential | Oxidation Energy, Reduction Energy, Neutral Energy, IP, EA, … |
| Molecular Dynamic Simulation | Classic Molecular Dynamic Simulation | Diffusion coefficient, Density, Viscosity, Conductivity, Coordination Number, Mean Square Displacement, Radial Distribution Function, Mean Residence Time, Simulation Trajectory |

For QC calculations, CEMP uses the open access ORCA 6.0.1[17]. Each spreadsheet row defines a unique molecular task via the name and SMILES string of the molecules. For MD simulations, GROMACS 2018.8[18] is used based on users specified SMILES, number of molecules, temperature and simulation time. To ensure computational stability and reproducibility, advanced simulation parameters, such as time step, thermostat, and barostat settings are preoptimized without requirements for user adjustment. The platform supports batch submission and can automatically process large job queues.

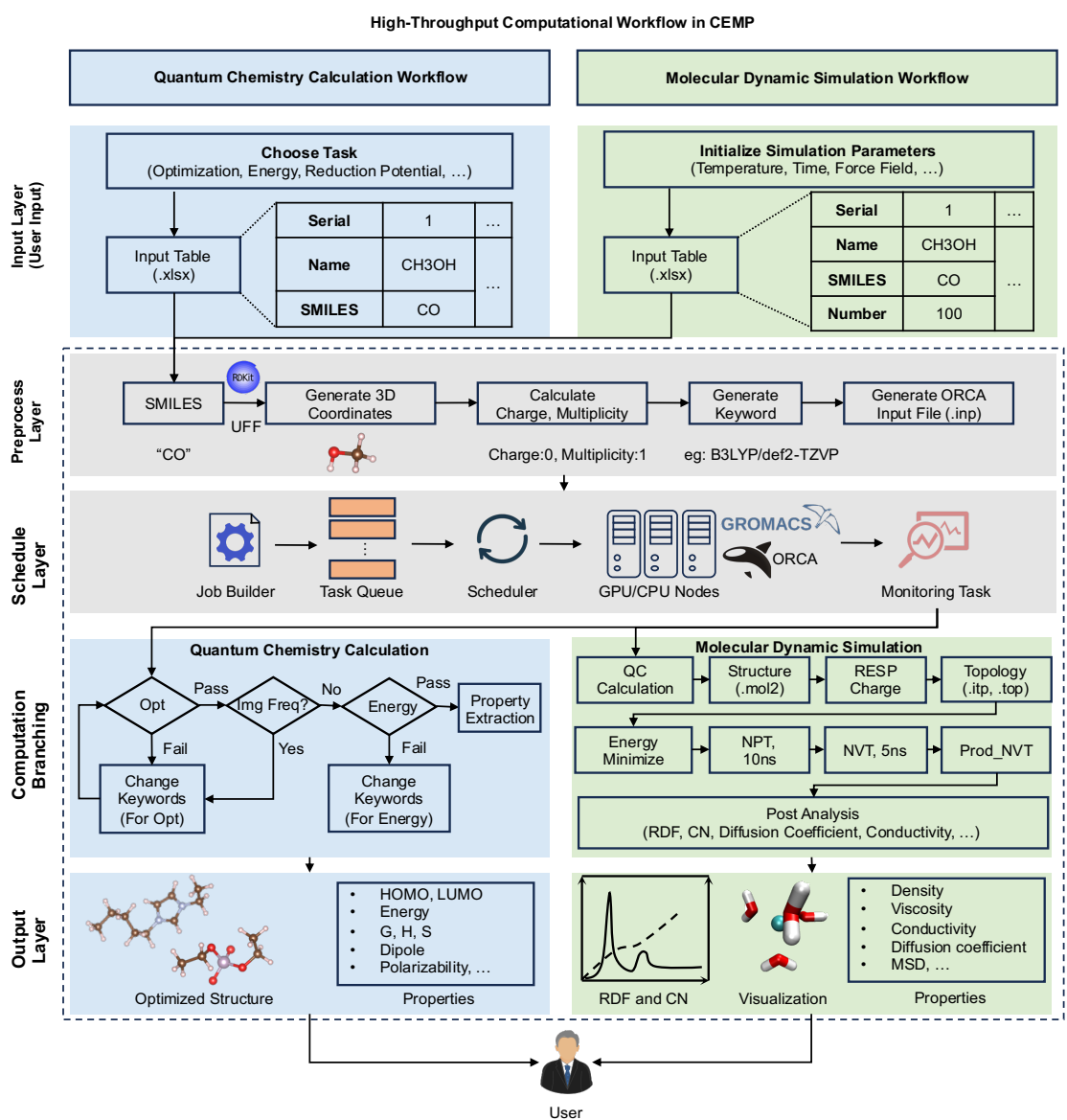

**Figure 3**. Schematic overview of the high-throughput automated computational workflow in the CEMP. The workflow consists of five vertically integrated layers, enabling end-to-end execution of quantum chemistry (blue) and molecular dynamics (green) tasks.

Upon submission, the preprocessing layer parses the spreadsheet and converts the SMILES strings into 3D geometries as illustrated in Methods section. Formal charges and multiplicities are assigned automatically, and input files are generated accordingly. The scheduling layer continuously monitors server load by dispatching CPU nodes for QC or GPU nodes for MD when slots are available. Currently up to three concurrent tasks are allowed to execute parallel with other submitted tasks added to a queue. During the computation stage, standard workflows are executed. The QC pipeline includes structure optimization, vibrational frequency analysis, and single-point energy calculation. At present, only preselected basis sets and functionals are supported as illustrated in Methods section, with plans to expand computational levels in future releases. The platform includes built-in error handling: convergence failures and imaginary frequencies are detected and corrected by adjusting initial wavefunctions or keywords to maximize task completion rates.

For MD simulations, GAFF[19] is used to assign force field parameters, with RESP charges[20] generated for each species, detailed parameter settings and computational protocols are provided in Methods section. Molecules are randomly packed into simulation boxes, followed by energy minimization and equilibration, leading into production runs. After simulation, a built-in post-processing module calculates key transport and coordination properties, such as diffusion coefficients and radial distribution functions. GAFF provides broad coverage for organic systems such as electrolytes and polymers, and additional force fields will be incorporated to support more diverse systems in future updates.

Upon completion, CEMP enters the results aggregation phase, automatically organizing output files and returning download links to users. QC tasks return key energy properties, optimized structures, and full process files, e.g. inputs, outputs and checkpoints. The outputs for MD tasks including quantum properties for each species, diffusion kinetics data, trajectory snapshots, visualized plots and raw data for custom plotting and analysis. All outputs are delivered in structured formats, including tabulated data and graphical representations to support interpretation and downstream analysis.

This high-throughput workflow provides a streamlined user experience with minimal input and maximal output, while maintaining computational robustness and full traceability. Its automation and scalability allow researchers to validate predictions and generate reliable data for training machine learning models.

## MULTI-SOURCE MATERIAL DATABASE FOR CLEAN ENERGY APPLICATIONS

As materials science shifts toward a data-driven paradigm, the creation of large-scale, high-quality and semantically rich databases has become essential to overcome longstanding bottlenecks in clean energy materials discovery. Existing platforms, however, remain dominated by first-principles data for inorganic crystals and therefore suffer from three shortcomings: 1) limited experimental data with available labels largely confined to electronic-structure quantities such as band gaps and total energies; 2) support for organic systems is minimal, even though practical applications rely on the composites material systems. To address these limitations, CEMP has developed a large-scale, FAIR-compliant (Findable, Accessible, Interoperable, Reusable) database that integrates four key material classes: small molecules, ionic liquids, polymers and crystals. CEMP systematically harmonizes data from experimental measurements, theoretical calculations and ML predictions using unified data structures and standardized pipelines.

All experimental records are annotated with source references to ensure traceability. All QC entries are labeled with the specific level of theory and calculation software used in each calculation, ensuring consistency and comparability across different compounds. ML-predicted values are accompanied by uncertainty metrics such as MAE and $R^2$, offering users confidence in the reliability of each prediction. All datasets are openly available in structured table format (.csv and .xlsx) for easy access and downstream use.

**Figure 4(a)** presents the overall scale of each CEMP sub-database and the origin of the contents. **Figure 4(b)** further disaggregates the distribution of data types across material systems. As experimental entries require manual curation and validation, they constitute the smallest portion of the dataset. In contrast, QC data expands steadily through routine platform updates. The majority of ML entries are derived from computationally generated structures, with their properties predicted using the trained ML models. The molecule database currently includes ~20,000 QC records. The ionic liquid database comprises ~8,000 QC entries, ~2,000 experimentally measured data points, and ~100,000 ML predictions. For polymers, the database contains ~10,000 QC entries, ~4,000 experimental records, and ~210,000 ML-predicted structures. The crystal database, which is under active development, includes ~10,000 ab-initio entries obtained from the Materials Project[7] and will be further expanded to encompass crystalline materials relevant to clean energy applications. **Figure 4(c)** highlights additional property types in the database beyond the standard quantum chemical descriptors listed in Table 2.

Table 2. Summary of standard quantum chemical properties available in the CEMP database.

| Property | Unit |
| --- | --- |
| Energy | Hartree |
| HOMO | eV |
| LUMO | eV |
| HOMO LUMO Gap | eV |
| Enthalpy | Hartree |
| Entropy | Hartree |
| Gibbs Free Energy | Hartree |
| Dipole | Debye |

To ensure data reliability and consistency, CEMP implements a rigorous quality control pipeline. Given the heterogeneous nature of experimental sources (peer-reviewed literature, handbooks, other databases), a robust statistical aggregation strategy is applied. For properties with multiple reported values, outliers are first identified and removed using the 1.5× IQR rule, and remaining values are aggregated using a 50% trimmed mean to ensure stability and robustness. However, we acknowledge that detailed experimental conditions are not always available, especially for properties that are sensitive to test protocols, such as in polymer systems. Addressing this challenge remains a key focus for future data expansion.

In the following sections, we outline the design objectives, key data types, and domain-specific use cases of the molecule, ionic liquid, and polymer databases. As the crystal database remains under active development, it is not included in the present discussion.

**Molecule Database**: The molecule database is designed to support molecular-level structure–property analysis for applications in electrolyte formulation and polymer precursor design. It focuses on organic species commonly used in battery electrolytes and polymer synthesis, including reported electrolytes[21], Li$^+$/Na$^+$/K$^+$ salts, and widely available monomers (sourced from the eMolecules). The database currently contains ~20,000 property entries and provides three core computed properties: metal ion-ligand binding energies, solvation free energies and redox property. As shown in **Figure 4(d)**, the binding energy distributions of Li$^+$, Na$^+$, and K$^+$ with common anions or solvent molecules follow the expected trend of Li$^+$ > Na$^+$ > K$^+$, consistent with their relative electronegativities and ionic radii. These data provide theoretical guidance for understanding ion solubility, coordination environments, and electrolyte stability. **Figure 4(e)** presents solvation free energy distributions of various monomers in common organic solvents. More negative values indicate better solubility, an essential consideration in selecting appropriate solvents for polymerization. The heatmap reveals clear trends in solute–solvent compatibility, offering insight into solvent polarity preferences that influence polymerization efficiency and product dispersity. **Figure 4(f)** compares the redox windows of monomers, computed using IP–EA methods in Methods section. The redox property is crucial for predicting the oxidative/reductive resilience of materials, especially in high-voltage battery environments.

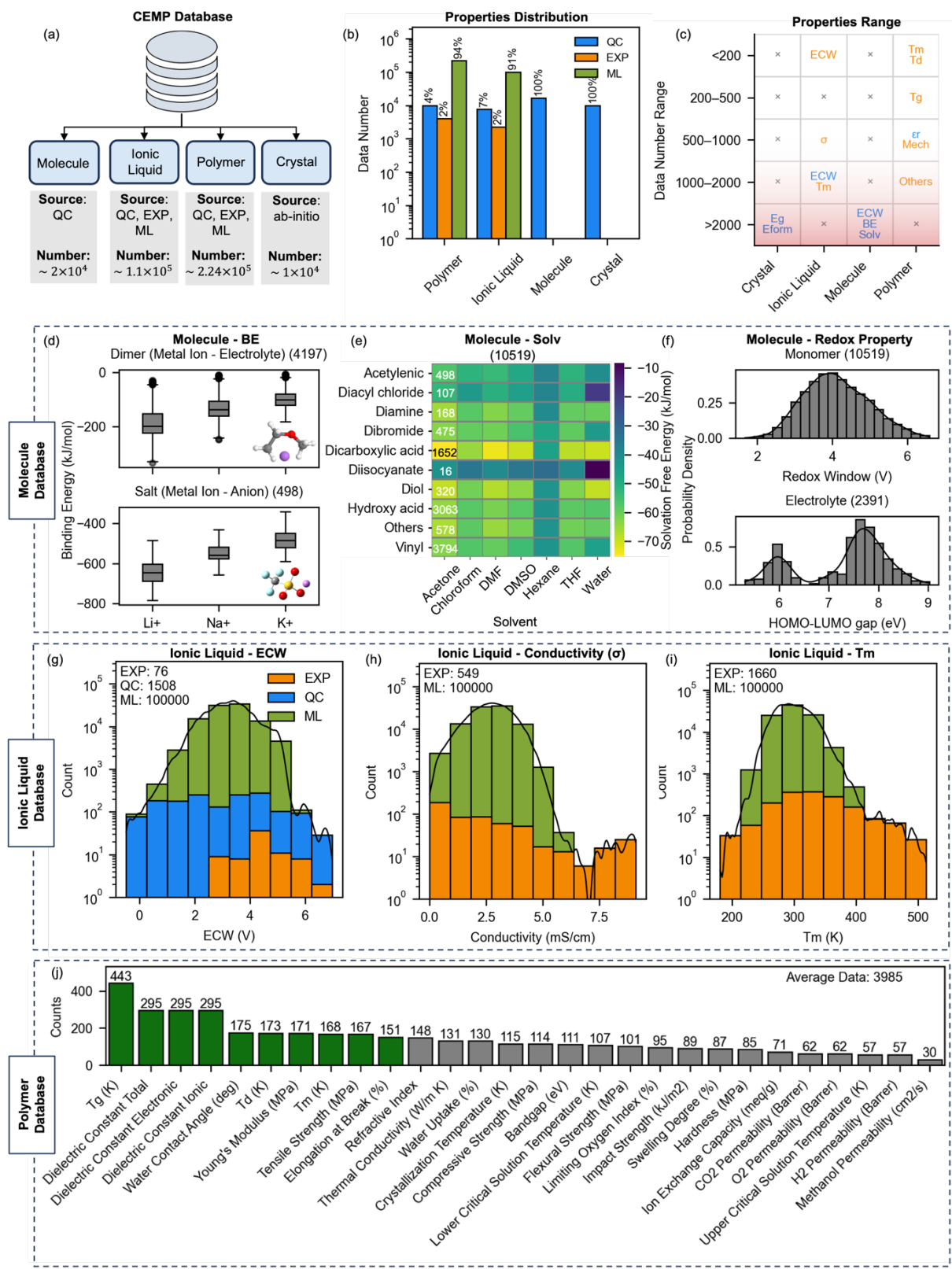

**Figure 4:** Overview of the CEMP database and its data composition. (a) Summary of database categories, including data sources and total entry counts. (b) Proportions and absolute numbers of data entries from different sources, including QC, ML, and experimental (EXP) across the three databases. (c) Distribution of property types and their data volumes in each database. (d)

Box plot showing the distribution of binding energies between alkali metal cations and electrolyte components or anions. (e) Heatmap of solvation free energies for monomers across various solvents; numerical labels indicate the number of monomers per solvent category. (f) Histogram comparing electrochemical windows (ECWs) of monomers and HOMO–LUMO gaps of corresponding electrolytes. (g) Distribution of ECWs derived from QC calculations and ML predictions. (h) Ionic conductivity distribution including both experimental measurements and ML predictions. (i) Melting point distribution based on experimental and ML-predicted data. (j) Bar chart summarizing the number of data points for key polymer properties, including experimental and QC-derived values. Numbers in parentheses indicate the total number of data points available for each property or category.

**Ionic Liquid Database**: The ionic liquid database supports the design and screening of ionic liquids for energy storage applications. It includes ~2,000 experimentally reported ILs[22] and over 100,000 algorithmically generated candidates, providing a broad structural and property space for large-scale screening. **Figure 4(g)** shows the distribution of electrochemical windows (ECWs) for ionic liquids, computed using IP–EA methods. These data help identify ionic liquids that are electrochemically stable across a wide voltage range, essential for compatibility with various anode/cathode materials, particularly in solid-state or high-voltage batteries. **Figure 4(h)** displays room-temperature ionic conductivity values, a key parameter for charge transport performance. **Figure 4(i)** summarizes the melting point distributions of ionic liquids, based on both experimental measurements and ML predictions. Melting point is a critical property that determines the operating temperature window and phase stability in practical devices.

**Polymer Database:** The polymer database provides a multi-scale data resource for the design of functional polymer materials. It focuses on homopolymer systems, combining curated structural and experimental data[23-26] with generated polymer candidates[27]. It is currently one of the most comprehensive datasets available for polymer structure–property modeling. **Figure 4(j)** shows a bar chart summarizing data coverage for 26 key polymer properties. These properties span thermodynamic, mechanical, and electronic domains, providing a rich foundation for multi-task learning, multi-objective optimization, and multi-scale modeling efforts.

All sub-databases are openly released under the CC BY–NC–SA 4.0 license, allowing free use, redistribution, and modification for non-commercial purposes, provided derivative works retain the same license. The platform is updated quarterly, incorporating newly published experimental data, high-throughput calculations, and ML-predicted labels.

**PROPERTY PREDICTION MODELS INTEGRATED IN CEMP**

Macroscopic material properties, such as mechanical strength and electrochemical stability, are essential for ensuring device compatibility and operational reliability. While CEMP already integrates an automated high-throughput computing workflow and a large-scale multi-source database, traditional first-principles methods remain computationally intensive and limited in chemical space coverage when applied to these complex and macroscopic properties. To address this challenge, CEMP incorporates a suite of ML property prediction models to enable fast and accurate performance estimation, complementing conventional simulation approaches. As shown in **Figure 5(a)**, 12 ML models have been deployed across four functional modules of the platform, tailored to different material systems and application scenarios.

Model selection is based on the characteristics of the prediction task and data structure. For small datasets with high label variance, such as experimental properties of ionic liquids and polymers, XGBoost models are employed due to their robustness to noise and ability to capture nonlinear relationships[28,29]. For crystalline materials, which naturally exhibit graph-like structures, graph neural network (GNN) architectures such as Graph Attention Networks (GAT) and Graph Convolutional Networks (GCN) are used. For the time-series prediction of battery discharge behavior under various operating conditions, a Transformer model is adopted to capture complex temporal and multi-variable interactions.

All models are trained on experimentally validated data or standardized QC results from the CEMP database, ensuring traceability and consistency across tasks. In terms of feature engineering, the Ionic Liquid and Polymer modules use SMILES strings as input and generate 2048-bit Morgan fingerprints[30] (radius = 2) via RDKit, which are fed into XGBoost or MLP models for property prediction (**Figure 5(b)**). The Crystal Module accepts CIF files as input, and uses Pymatgen[31] to extract graph-based structural features, including atom types (nodes), bond indices (edges) and edge features. To enable self-supervised learning of crystal representations, a momentum contrastive learning (MOCO)[32] strategy is implemented. This approach facilitates the extraction of generalizable features from unlabeled data and improves performance in low-data regimes. The workflow, which remains under active development, includes transfer learning for downstream property prediction. Four model variants (MOCO+GAT, MOCO+GCN, GAT, and GCN) are benchmarked and the best-performing model is selected through ensemble voting. The model architecture is illustrated in **Figure 5(c)**. For C-rate curve prediction to evaluate the battery cycling performance, internal battery recipe and external experimental conditions, including cathode type, electrolyte composition, electrolyte thickness and temperature are encoded as feature vectors into a Transformer encoder, enabling the prediction of voltage vs. capacity at varying C-rate across cycles (**Figure 5(d)**). Details of model training procedures and hyperparameter optimization are provided in as Methods section. Additional training protocols and model-specific configurations will be made available in forthcoming publications.

**Figure 5(e)** summarizes the performance of all models on independent test sets, evaluated using MAE and $R^2$. All models achieve promising $R^2$ values between 0.64 and 0.94. Notably, the C-rate model performs best in capturing dynamic behavior ($R^2 = 0.94$). **Figure 5(f)** showcases highly consistent results for representative materials, including the conductivity, ECW and melting point of $C_2mimFSI$ (ionic liquid), the electrochemical properties of P2–$NaNiO_2$ (cathode material), and the thermal, mechanical, and dielectric properties of PPO (polymer; $T_g = 400.29$ K, $E = 1356$ MPa, $T_m = 452$ K, $\varepsilon_r = 3.31$). **Figure 5(g)** demonstrates the platform's full prediction capability for C-rate discharge curves, including time-resolved voltage and capacity profiles over multiple cycles, highlighting its potential for real-world battery behavior modeling.

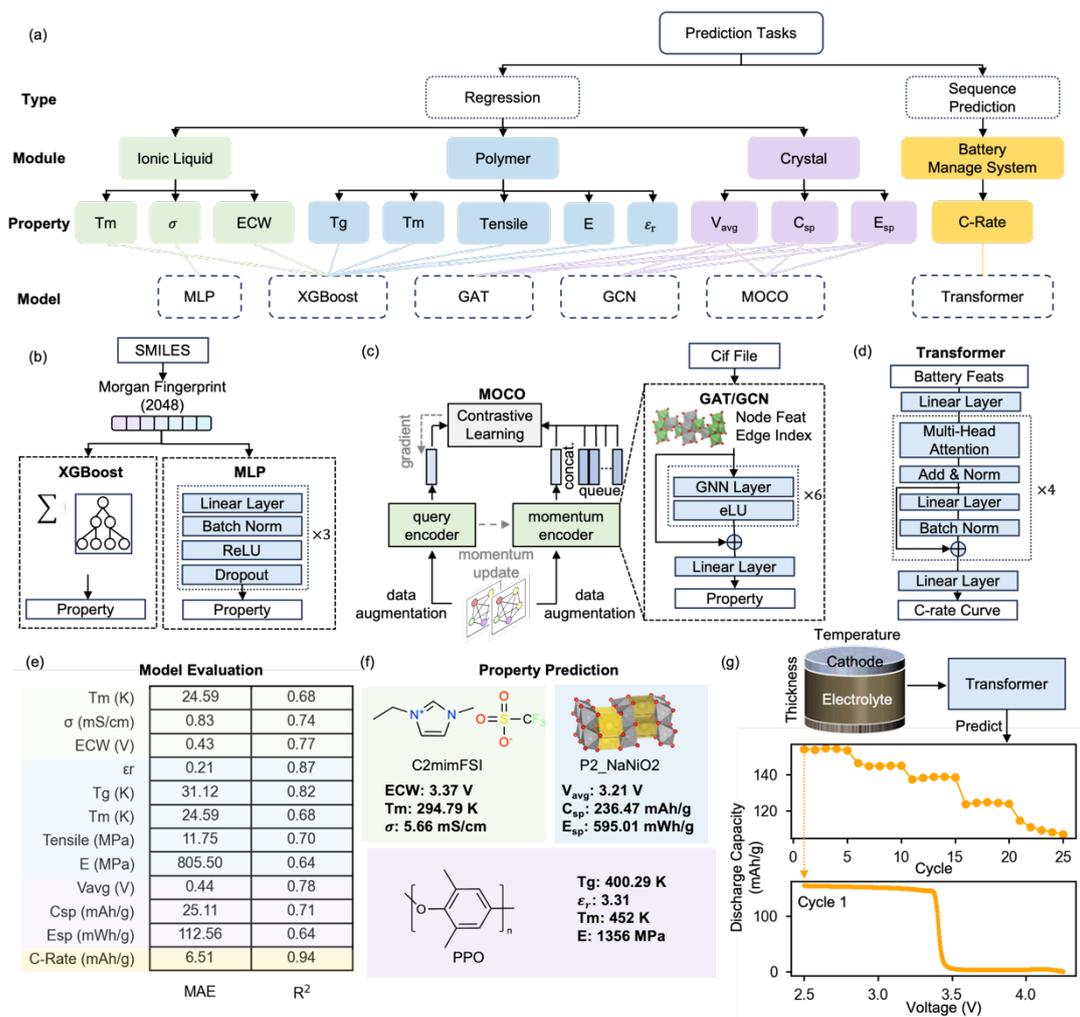

**Figure 5.** Overview of property prediction tasks and model architectures integrated into the CEMP. (a) CEMP supports a diverse set of prediction tasks across four major modules. Target properties include melting point (Tm), ionic conductivity (σ), ECW, glass transition temperature (Tg), tensile strength, Young's modulus (E), dielectric constant ($\varepsilon_r$), average voltage ($V_{avg}$), specific capacity ($C_{sp}$), specific energy ($E_{sp}$), and C-rate discharge profiles. (b) Model architecture of XGBoost and MLP models. (c) Model architecture of MOCO, GCN and GAT. (d) Model architecture of Transformer. (e) Model performance across all prediction tasks, evaluated using independent test sets. Each colored block represents a different module. (f) Representative prediction examples. (g) Predicted battery C-rate performance curves using the Transformer model.

**FUTURE OUTLOOK FOR CEMP**

The current version of CEMP integrates high-throughput computing, a structured multi-source database, and ML models for macroscopic property prediction, laying the foundation for a closed-loop workflow from data acquisition to model validation. As user demand grows and task complexity increases, future developments will focus on improving scalability, interactivity and computational efficiency, as shown in **Figure 6**. In the near future, efforts will focus on AI-accelerated simulations, including machine learning-based structure optimization, ML force fields, and rapid property predictors to significantly reduce computational overhead and improve throughput. In the mid-term, the platform will introduce intelligent agents and a Python package (CEMP-Py) to enable personalized workflows, and seamless integration into user-defined pipelines. Looking ahead, CEMP will transition to a cloud-native, GPU-accelerated architecture with elastic resource scheduling, enabling high-throughput, parallel computing at scale.

By integrating multimodal data, modular AI models and high-throughput computation workflows, CEMP aspires to become a fully automated, intelligent and extensible platform that accelerates the discovery of clean energy materials.

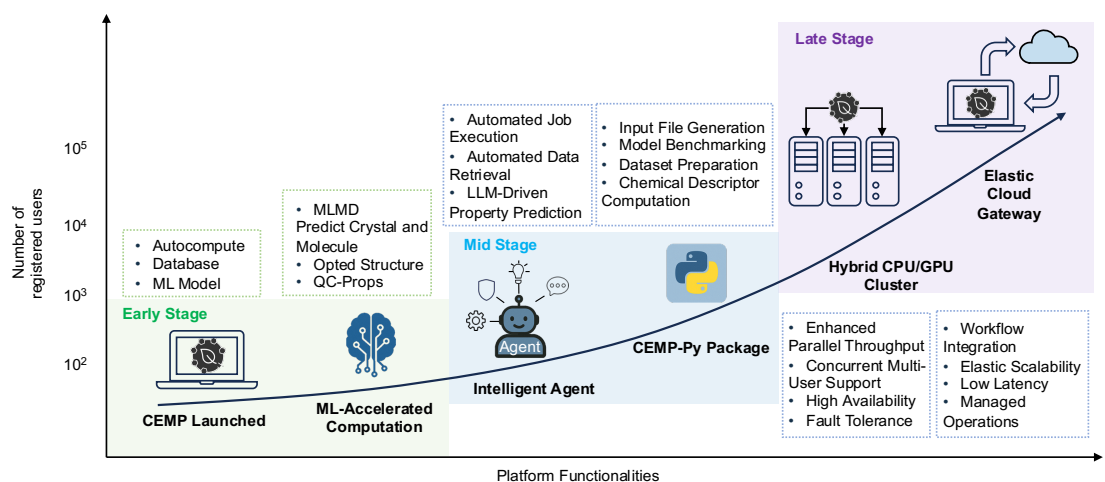

**Figure 6.** Evolution of the CEMP in functionality and user scale. Early stage: Initial deployment with core capabilities, including online high-throughput computation, database, and property prediction models. Mid stage: Introduction of intelligent agents and the Python package (CEMP-Py), enabling automated task execution, data retrieval, and LLM-assisted property prediction. Late stage: Full deployment on hybrid CPU/GPU clusters with elastic

cloud infrastructure, supporting large-scale concurrent computation and multi-user scheduling.

**CONCLUSION**

CEMP is designed to address the growing need for a paradigm shift in materials science from empirical intuition to data-driven and algorithm-oriented research paradigm. Focusing on clean energy applications, the platform integrates high-throughput computational workflows, a large-scale heterogeneous database and multi-scale ML models, covering four major material classes: small molecules, ionic liquids, polymers, and crystals. By adhering to FAIR principles and adopting modular architecture, CEMP establishes a unified framework linking structure, property and performance, thereby enhancing the efficiency and scalability of materials design. The platform supports automated simulations from SMILES or structural files and leverages AI models for rapid property prediction, enabling end-to-end workflows for screening, optimization and validation. CEMP incorporates specialized model architectures across material types, demonstrating strong generalization and task flexibility. Looking forward, the platform will continue evolving with the integration of intelligent agents and cloud-native infrastructure to further enhance usability. We envision CEMP not only as an efficient research tool, but as a foundational infrastructure that will help accelerate the discovery of advanced materials across clean energy domains.

**METHODS**

**SMILES-to-3D Structure Conversion Algorithm**

The initial three-dimensional structure of a target compound was generated from its SMILES representation using OpenBabel[33]. First, the SMILES string was input into OpenBabel, which automatically constructed a preliminary 3D structure. Subsequently, a conformational search was performed to generate 50 distinct conformers. Each conformer was then subjected to energy minimization using the Universal Force Field (UFF)[34], and the corresponding minimized energy values were recorded. The conformer with the lowest energy was selected as the final 3D structure of the compound. For validation, this approach was compared to an alternative method utilizing RDKit in conjunction with the MMFF94 force field[35]. The results demonstrated that, particularly for inorganic compounds (e.g., $PF_6^-$, $BF_4^-$), the OpenBabel-based method with conformational search and UFF yielded more reliable 3D structures than the RDKit/MMFF94 combination.

**Quantum Chemistry Calculation Details**

Quantum chemistry calculations were performed using ORCA 6.0.1[17]. Initial geometry optimizations and frequency calculations were carried out at the B3LYP[36]/def2-TZVP[37] level. Frequency analyses confirmed that each optimized geometry corresponded to a true minimum on the potential energy surface, as evidenced by the absence of imaginary frequencies, with DFT-D3 dispersion corrections applied throughout[38]. Subsequently, single point energy calculations were executed on these optimized structures at the wB97M-V[39]/def2-TZVP level.

**Molecular Dynamic Simulation Details**

Molecular Dynamic simulations were performed using Gromacs2018.8[18]. The simulations employed the General Amber Force Field (GAFF) for organic compounds[19], the OPC3 water model[40], and the Merz force field for metal ions[41]. Atomic charges were computed using Multiwfn package[42] via the restrained electrostatic potential (RESP) method. To account for charge transfer and polarizability effects in the condensed phase, all ionic RESP charges were uniformly scaled by a factor of 0.7 [43]. Periodic boundary conditions were applied in all three dimensions. Van der Waals and short-range electrostatic interactions were calculated with a cutoff of 1.2 nm. The equations of motion were integrated using the leap-frog algorithm with a time step of 2 fs. Temperature was maintained using a velocity-rescale thermostat[44], and pressure was controlled via a Berendsen barostat[45]. Coulomb interactions were treated with the Smooth Particle Mesh Ewald (SPME) method[46], with temperature and pressure coupling time constants set to 0.5 ps and 2 ps, respectively. All bonds were constrained using the LINCS algorithm[47]. Prior to production runs, the system underwent a rigorous pre-equilibration protocol consisting of energy minimization, followed by a 5 ns NPT simulation and a 5 ns annealing NVT simulation (from 400 K to the target temperature) to ensure equilibrium.

**Quantum Chemistry Calculation of Redox Potentials**

Redox potentials were computed using the ionization potential–electron affinity (IP–EA) method, which has been shown to provide an optimal balance of predictive accuracy and computational efficiency[48]. This method focuses on the transition of gas-phase ions from the neutral state to their oxidized and reduced forms. The redox potentials were calculated according to the following equations:

$$E_{red} = E(N-1) - E(N) \qquad (1)$$

$$E_{ox} = E(N) - E(N+1) \qquad (2)$$

Here, $E(N)$ denotes the ground state energy of the molecule, while $E(N+1)$ and $E(N-1)$ represent the energies of the reduced and oxidized states, respectively. The energy evaluation was performed using the same protocol as that employed for single point energy calculations.

**Ionic Liquid Property Prediction Model**

To predict the properties of ionic liquids, three independent XGBoost models were developed for estimating room-temperature conductivity, ECW, and melting point. The dataset, collected from ILthermo[22], was partitioned into training, validation, and test sets in an 8:1:1 ratio. For each IL, SMILES representations of ionic liquids were used to compute Morgan fingerprints[30] (radius = 2) each. Given the differing feature requirements for each target property, separate models were trained for conductivity, ECW, and melting point.

**Crystal Property Prediction Model**

For crystal property prediction, three graph neural network (GNN) models, including GAT, GCN and a MOCO+GAT were constructed. This framework predicts key properties including average voltage, gravimetric capacity, and gravimetric energy density from CIF files uploaded by the user. Initially, the CIF file is converted into a structure object using pymatgen[31]. For each crystal sample, an encoding strategy is employed where the elemental type of each atom serves as its node feature. The local environment of each atom is defined by all atoms within a 5 Å radius, and the interatomic distances within this neighborhood are expanded using Gaussian basis functions to serve as edge features. These node and edge features are then input into the GNN models. After multiple layers of feature extraction, a 64-dimensional structural representation is obtained, which is processed through a regression head to yield the predicted crystal properties.

The dataset for the crystal models was sourced from the Materials Project[7] and partitioned into training, validation, and test sets in an 8:1:1 ratio. In the case of the MOCO+GAT model, data augmentation was performed by randomly omitting 1–5% of the node or edge features to generate similar samples.

**Structural Visualization**

All structural visualizations in the CEMP were performed using the JSmol plugin[49], which facilitates interactive three-dimensional representations of molecular and crystalline structures.

**AUTHOR CONTRIBUTIONS**

J.W. led the full-stack and algorithmic development of the Autocompute, Ionic Liquid, Polymer, and Crystal modules. He was also responsible for data collection, high-throughput computational workflows, and the initial drafting and subsequent revision of the manuscript. J.J. developed both the full-stack infrastructure and algorithmic framework for the Battery Management System module. Y.W. conceived and supervised the overall project, and contributed to the manuscript editing and refinement.

**CONFLICT OF INTERESTS**

The authors declare no competing financial interest.

**DATA AVAILABILITY**

The CEMP platform is publicly accessible at **https://cleanenergymaterials.cn**, where users can explore the database, submit calculation tasks and interact with deployed ML models. Please note that the source code for the platform architecture and high-throughput computational workflows is currently not open to the public.

**Table 3.** Access links to core functional modules in the CEMP.

| Module | Functionality | Access URL |
|---|---|---|
| Autocompute | MD workflow | https://cleanenergymaterials.cn/autocompute/MDCompute |
| | QC workflow | https://cleanenergymaterials.cn/autocompute/QCcompute |
| | Small molecule database | https://cleanenergymaterials.cn/autocompute/Database |
| Ionic Liquid | IL property prediction | https://cleanenergymaterials.cn/ionic_liquid/XGBoost_predict |
| | IL database | https://cleanenergymaterials.cn/ionic_liquid/ionic_liquid_database_card_page |
| Polymer | Polymer database | https://cleanenergymaterials.cn/polymer/polymer_database |
| | Polymer generator | https://cleanenergymaterials.cn/polymer/generate_polymer_display |
| | Polymer property prediction | https://cleanenergymaterials.cn/polymer/polymer_predict_page |
| Cystal | Cystal database | https://cleanenergymaterials.cn/crystals/crystal_selected/ |
| | Cystal property prediction | https://cleanenergymaterials.cn/crystals/crystal_prediction/ |
| Battery Manage System | Short-Term Prediction | https://cleanenergymaterials.cn/battery_manage_system/short_prediction/ |

**ACKNOWLEDGMENTS**

We appreciate the support of the National Natural Science Foundation of China (92372126, 52373203), the Excellent Young Scientists Fund Program and the AI for Science Foundation of Fudan University (FudanX24AI014).